\documentclass[12pt]{article}

\usepackage{amsmath,amsthm,amssymb,amscd}

\usepackage{graphicx, float, epsfig}
\usepackage[hidelinks]{hyperref}
\usepackage{harvard}
\usepackage{rotating}
\usepackage{pww}

\usepackage{enumitem} 
\usepackage{caption, subcaption} 
\usepackage{capt-of} 
\usepackage[flushleft]{threeparttable} 
\usepackage{multirow}
\usepackage{booktabs}

\usepackage{xcolor,soul} 
\usepackage[linesnumbered,ruled,vlined]{algorithm2e} 

\usepackage{authblk}

\newtheorem{theorem}{Theorem} 

\title{\bf Non-crossing convex quantile regression}

\author[a]{Sheng Dai}
\author[a]{Timo Kuosmanen}
\author[b,\footnote{Corresponding author. \newline
\hspace*{5mm} \textit{E-mail addresses:} \texttt{sheng.dai@aalto.fi (S. Dai)}, \texttt{timo.kuosmanen@aalto.fi (T. Kuosmanen)},\\
\hspace*{34mm} \texttt{xun.zhou@york.ac.uk (X. Zhou)}.}]{Xun Zhou}
\affil[a~]{Aalto University School of Business, 02150 Espoo, Finland}
\affil[b~]{Department of Environment and Geography, University of York, York YO10 5NG, UK}
\date{April 2022}

\begin{document}
\captionsetup[figure]{labelfont={bf},labelformat={default},labelsep=period,name={Fig.}}
\captionsetup[table]{labelfont={bf},labelformat={default},labelsep=period,name={Table}}

\citationmode{abbr}
\bibliographystyle{jbes}

\maketitle

\vfill
\vfill

\begin{abstract}
\noindent
  Quantile crossing is a common phenomenon in shape constrained nonparametric
  quantile regression. A recent study by \citeasnoun{Wang2014c} has proposed to address this problem by imposing non-crossing constraints to convex quantile regression. However, the non-crossing constraints may violate an intrinsic quantile property. This paper proposes a penalized convex quantile regression approach that can circumvent quantile crossing while better maintaining the quantile property. A Monte Carlo study demonstrates the superiority of the proposed penalized approach in addressing the quantile crossing problem.
\\[5mm]
\end{abstract}
\vfill

\thispagestyle{empty}
\newpage
\setcounter{page}{1}
\setcounter{footnote}{0}
\pagenumbering{arabic}
\baselineskip 20pt

%

\section{Introduction}\label{sec:intro}

Quantile estimation has been widely applied in various fields of economics and econometrics (see, e.g.,
\citename{Wang2014c}, \citeyear*{Wang2014c}; \citename{Jradi2019},
\citeyear*{Jradi2019}; \citename{Tsionas2020b}, \citeyear*{Tsionas2020b};
\citename{Kuosmanen2020b}, \citeyear*{Kuosmanen2020b}; \citename{Zhao2021},
\citeyear*{Zhao2021}). However, when multiple quantiles are separately estimated
to obtain a family of conditional quantile functions, two or more quantile
curves may cross on the condition that the distribution functions and their
associated inverse functions are not monotone increasing (\citename{He1997},
\citeyear*{He1997}). Such quantile crossing is a longstanding problem in quantile regression.

To our knowledge, there are three commonly seen approaches to avoid quantile crossing: post-processing, stepwise estimation, and simultaneous estimation. In the post-processing procedure, a non-crossing assumption is usually enforced via a sorting or monotonic rearrangement of the original estimated non-monotone functions (e.g., \citename{Dette2008}, \citeyear*{Dette2008}; \citename{Chernozhukov2010}, \citeyear*{Chernozhukov2010}). This indirect approach is effective in estimating the conditional quantile, but lacks the ability to quantify the effects of the predictors (\citename{Bondell2010}, \citeyear*{Bondell2010}). In the stepwise procedure, it prevents an estimated quantile function from crossing the previously estimated one by adding an extra set of non-crossing constraints iteratively to the regression model (e.g., \citename{Wu2009a}, \citeyear*{Wu2009a}); but this approach cannot offer simultaneous estimates. In the simultaneous estimation, non-crossing constraints are imposed to ensure that the estimated conditional quantile functions are monotone nondecreasing, with all quantiles being estimated simultaneously (e.g., \citename{Takeuchi2006}, \citeyear*{Takeuchi2006}; \citename{Bondell2010}, \citeyear*{Bondell2010}). More recently, \citeasnoun{Wang2014c} extend this simultaneous estimation technique to convex quantile regression (sCQR). However, the non-crossing constraints may disturb the quantile property (\citename{Takeuchi2006}, \citeyear*{Takeuchi2006}).

This paper develops a new non-crossing approach to nonparametric quantile function estimation. Compared with sCQR (\citename{Wang2014c}, \citeyear*{Wang2014c}), our approach based on the penalized convex quantile regression (pCQR) independently estimates multiple non-crossing quantiles but can better satisfy the intrinsic quantile property. Furthermore, the proposed pCQR approach can fit the true quantile functions more reliably and robustly.

%

\section{Penalized convex quantile regression}\label{sec:pcqr}

Consider a general nonparametric regression model with observations
$\{\bx_i, y_i \}_{i=1}^n$ satisfying
\begin{equation}
	\begin{aligned}
		\label{eq:reg}
    y_i = f(\bx_i) + \varepsilon_i, \qquad \mbox{for } i = 1,\ldots, n,
	\end{aligned}
\end{equation}	
where $y_i \in \real$ and $\bx_i \in \real^d$ are output and inputs variables, and $\varepsilon_i$ is a error term with zero mean. Accordingly, for a given quantile $\tau \in (0,1)$, the nonparametric quantile function $Q_y(\tau \,|\, \bx)$ is defined as
\begin{equation}
	\begin{aligned}
		\label{eq:qreg}
		Q_{y_i}(\tau \, | \, \bx_i)=f(\bx_i)+F_{\varepsilon_i}^{-1}(\tau),
	\end{aligned}
\end{equation}	
where $F_{\varepsilon_i}$ is the distribution function of the error term
$\varepsilon_i$.

To estimate quantiles empirically, we resort to convex quantile regression (CQR) that
does not require any assumptions about the functional form of the regression function $f$ or its smoothness, but imposes the shape constraints such as monotonicity and
concavity. Specifically, CQR estimates the quantile function \eqref{eq:qreg} by solving the following linear programming problem (\citename{Wang2014c}, \citeyear*{Wang2014c})
\begin{alignat}{2}
	\underset{\alpha,\bbeta,\varepsilon^\text{+},\varepsilon^{-}}{\mathop{\min }}&\,\tau \sum\limits_{i=1}^{n}{\varepsilon _{i}^{+}}+(1-\tau )\sum\limits_{i=1}^{n}{\varepsilon _{i}^{-}}  &{}&  \label{eq:cqr}\\ 
	\mbox{\textit{s.t.}}\quad
	& y_i=\mathbf{\alpha}_i+ \bbeta_i^{'}\bx_i+\varepsilon_i^{+}-\varepsilon_i^{-} &\quad& \forall i \notag \\
	& \alpha_i+\bbeta_i^{'}\bx_i \le \alpha_h+\bbeta_h^{'}\bx_i  &{}& \forall i,h \notag \\
	& \bbeta_i\ge \bzero &{}& \forall i  \notag \\
	& \varepsilon_i^{+}\ge 0,\ \varepsilon_i^{-} \ge 0 &{}& \forall i \notag
\end{alignat}
where the first set of constraints can be interpreted as a multivariate regression equation, the second set of constraints imposes concavity on the quantile function, the third set of constraints guarantees monotonicity, and the last refers to sign constraints of the error terms. Note that there exists an intrinsic quantile property in terms of the optimal solutions to problem \eqref{eq:cqr},  $\hat{\varepsilon}_i^{+} $and $\hat{\varepsilon}_i^{-}$.

\begin{theorem}
  For any $\tau \in (0, 1)$, the number of strict positive residuals ($\hat{\varepsilon}_i^{+}>0$) by $n_\tau^+$ and the number of strict negative residuals ($\hat{\varepsilon}_i^{-} > 0$) by $n_\tau^-$ always satisfy the inequalities:
\[
\frac{n_\tau^+}{n} \le 1-\tau \quad \mbox{and} \quad \frac{n_\tau^-}{n} \le \tau.
\]
\label{the:the1}
\end{theorem}\vspace{-2em}
\begin{proof}
See proofs in \citeasnoun{Wang2014c} and \citeasnoun{Kuosmanen2020b}.
\end{proof}

Compared with the conventional full frontier estimation, the quantile function estimation is more robust to random noise, heteroscedasticity, and the choice of direction vectors. However, when separately estimating each conditional quantile function $Q_y(\tau \,|\, \bx)$, CQR is likely to violate the assumption that the distribution functions and their associated inverse functions should be monotone nondecreasing; see Fig.~\ref{fig1} for an example of the quantile crossing problem detected in our empirical application of CQR.

\begin{figure}[H]
	\centering
	\begin{subfigure}[b]{1\textwidth}
		\centering
		\includegraphics[width=0.8\textwidth]{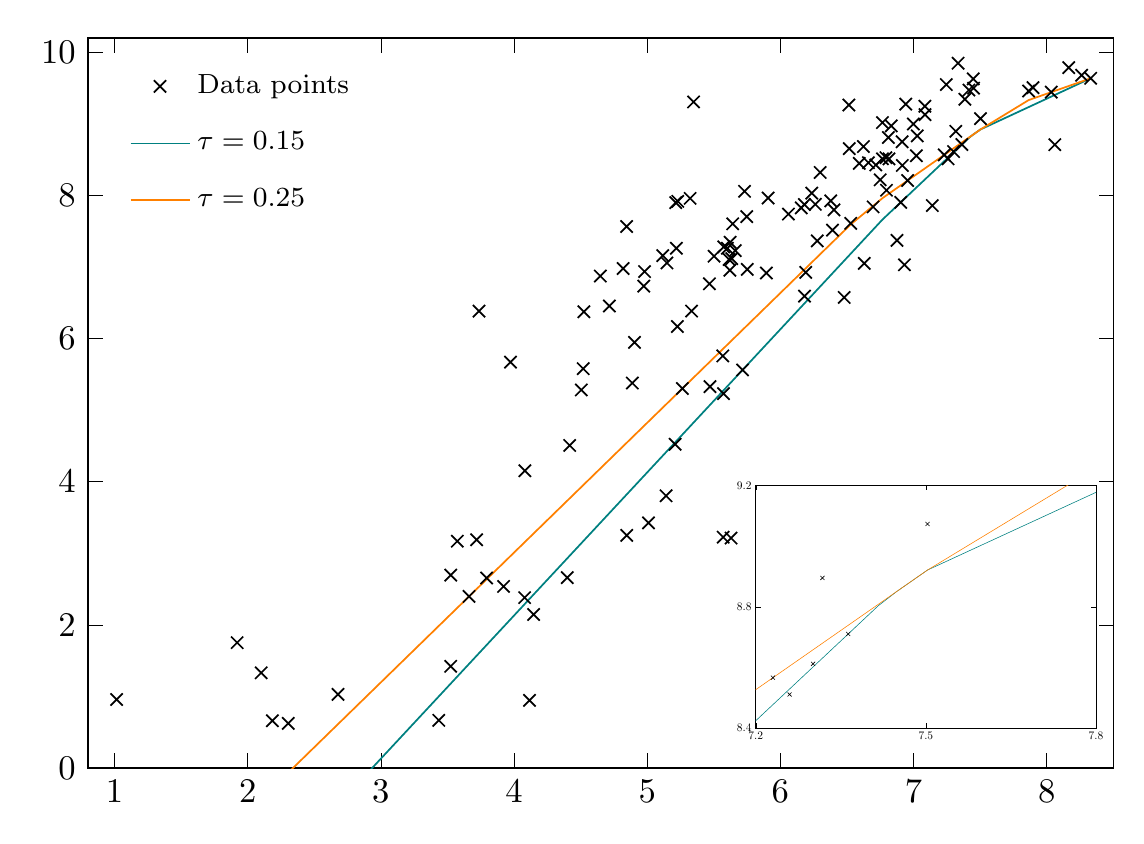}
		\caption[]%
		{{\small sCQR ($C=0$) }}    
		\label{fig2:a}
	\end{subfigure}
	\begin{subfigure}[b]{1\textwidth}  
		\centering 
		\includegraphics[width=0.8\textwidth]{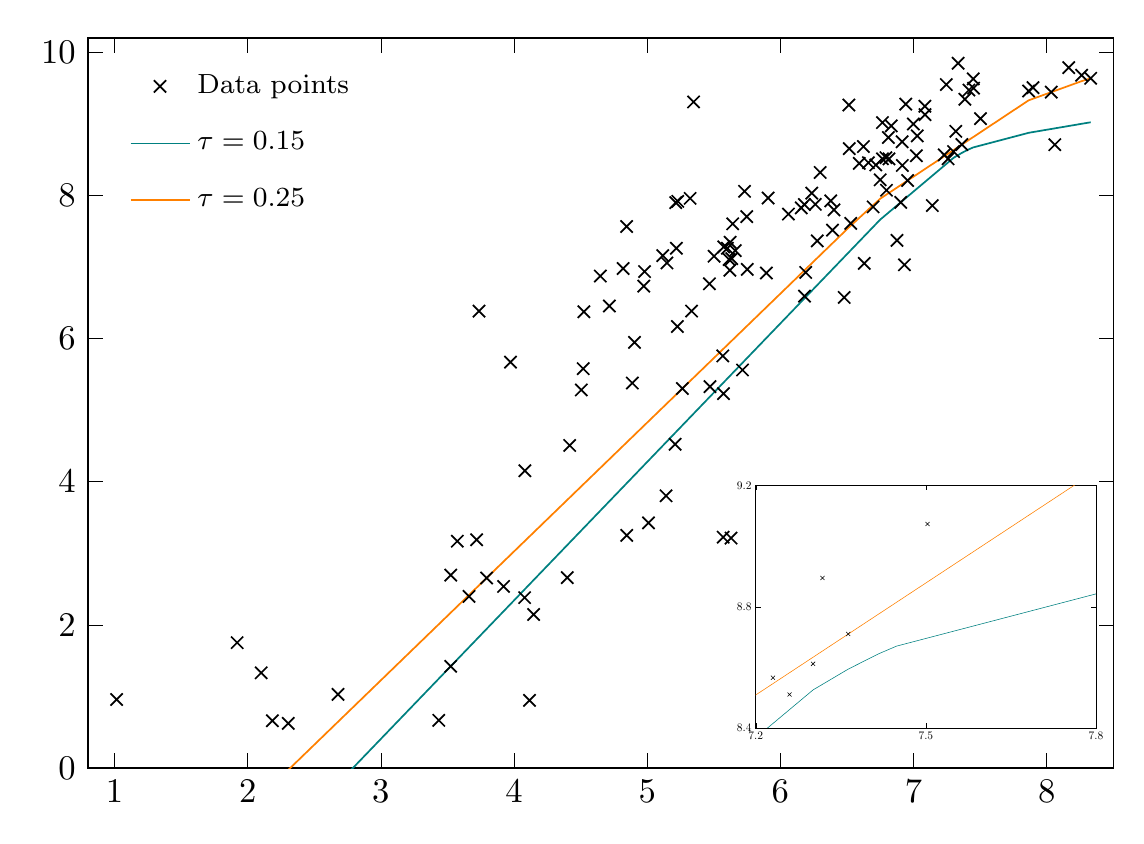}
		\caption[]%
		{{\small pCQR ($\gamma^* =0.01$)}}    
		\label{fig2:b}
	\end{subfigure}
	\caption[]%
	{\small Empirical illustration of estimated non-crossing quantile functions using the U.S. power plant data.} 
	\label{fig2}
\end{figure}

We notice that the quantile crossing problem could be addressed by simultaneous
estimation, which imposes an extra set of linear non-crossing constraints in the
CQR approach (see, e.g., \citename{Takeuchi2006}, \citeyear*{Takeuchi2006};
\citename{Wang2014c}, \citeyear*{Wang2014c}). Following \citeasnoun{Wang2014c}, the
simultaneous convex quantile regression (sCQR) estimator of $j$ conditional
quantile functions at $0 < \tau_1<\tau_2<\cdots<\tau_j<1$ is formulated as
\begin{alignat}{2}
	\underset{\alpha,\bbeta,\varepsilon^\text{+},\varepsilon^{-}, C}{\mathop{\min }}&\, \sum\limits_{j=1}^{\Js} \Big( \tau_j \sum\limits_{i=1}^{n}{\varepsilon_{i,j}^{+}}+ (1-\tau_j) \sum\limits_{i=1}^{n}{\varepsilon_{i,j}^{-}} \Big) &{}&  \label{eq:scqr}\\ 
	\mbox{\textit{s.t.}}\quad
	& y_{i,j}=\mathbf{\alpha}_{i,j}+ \bbeta_{i,j}^{'}\bx_i+\varepsilon_{i,j}^{+}-\varepsilon_{i,j}^{-} &\quad& \forall i, j \notag \\
	& \alpha_{i,j}+\bbeta_{i,j}^{'}\bx_i \le \alpha_{h,j}+\bbeta_{h,j}^{'}\bx_i  &{}& \forall i, h, j \notag \\
	& \alpha_{i,j} + \bbeta_{i,j}^{'}\bx_i + C_{i,j} \le \alpha_{i,j+1} + \bbeta _{i,j+1}^{'}\bx_i  &{}& \forall i, 1\le j \le \Js-1 \notag \\
	& \bbeta_{i,j}\ge \bzero &{}& \forall i, j  \notag \\
	& \varepsilon_{i,j}^{+}\ge 0,\ \varepsilon_{i,j}^{-} \ge 0 &{}& \forall i, j \notag
\end{alignat}
where $C_{i,j} \ge 0$ are small nonnegative constants for quantiles, which are introduced in sCQR to ensure that $Q_y(\tau_j \,|\, \bx) \le Q_y(\tau_{j+1} \,|\, \bx), \forall i \text{ } \text{and} \text{ } j \in \Js$. For the purpose of non-crossing, $C$ can simply be given by zero; that is, there may exist touching rather than crossing between two neighboring quantiles (see Fig.~\ref{fig2:a} for an illustration). In practice, however, after enforcing the non-crossing constraints, sCQR may violate the quantile property (Theorem \ref{the:the1}) due to the fact that the approach simultaneously optimizes for both the quantile property and the non-crossing property (\citename{Takeuchi2006}, \citeyear*{Takeuchi2006}).

This paper proposes an alternative to sCQR to address the quantile
crossing problem. By using the $L_2$-norm regularization on subgradients
$\bbeta_i$, we formulate penalized convex quantile regression (pCQR) as
\vspace{-1em}
\begin{alignat}{2}
	\underset{\alpha,\bbeta, \varepsilon^{+},\varepsilon^{-}}{\mathop{\min }}&\, \tau \sum\limits_{i=1}^{n}{\varepsilon _{i}^{+}}+(1-\tau )\sum\limits_{i=1}^{n}{\varepsilon _{i}^{-}} + \gamma\sum\limits_{i=1}^{n}||\bbeta_i||^2_2 &{}& \label{eq:pcqr}\\ 
	\textit{s.t.}\quad
	& y_i=\alpha_i+ \bbeta_i^{'}\bx_i+\varepsilon_i^+ -\varepsilon_i^- &\quad& \forall i \notag \\
	& \alpha_i+\bbeta_i^{'}\bx_i \le \alpha_h + \bbeta_h^{'}\bx_i &{}& \forall i,h \notag \\
	& \bbeta_i\ge \bzero &{}& \forall i \notag \\
	& \varepsilon _i^{+}\ge 0,\ \varepsilon_i^{-} \ge 0 &{}& \forall i \notag
\end{alignat}
where $\gamma \ge 0$ is the tuning parameter and $||\cdot||_2$ denotes the standard
Euclidean norm. As $\gamma$ approaches to zero, pCQR \eqref{eq:pcqr} collapses
to the original CQR problem \eqref{eq:cqr}. The rationale behind the avoidance of quantile crossing in pCQR lies in that as $\gamma \xrightarrow{} \infty$,
the regularization will dominate the minimization and then all estimated
subgradients $\bbeta_i$ ``flatten out" to 0. In this case, the estimated quantile functions will be horizontal lines (for $d=1$) or planes ($d>1$).

As $\gamma$ increases, the quantile property may also be disturbed in pCQR. But for sufficiently small $\gamma$, the quantile property
can be guaranteed in theory. Therefore, we design the following Algorithm
\ref{alg1} to obtain the supremum $\gamma^*$ so as to avoid quantile crossing and
ensure the quantile property as well as possible in estimating multiple quantile
functions. \vspace{0.5em}

\begin{algorithm}[H]\label{alg1} 
  \KwData{$\{\bx_i, y_i\}_{i=1}^n \in \real^d \times \real$, $\tau_1$ and $\tau_2$ ($\tau_1 < \tau_2$)}
  $\mathrm{out} = 0$ and $\gamma = 0$\;
  \While{$\mathrm{out} = 0$}{
  	Solve problem \eqref{eq:pcqr} with quantiles $\tau_1$ and $\tau_2$, separately, to calculate $\hat{Q}_y(\tau_1\,|\,\bx_i)$ and $\hat{Q}_y(\tau_2\,|\,\bx_i)$\;
    \eIf{$\sum\limits_{i=1}^{n}\mathbf{1}_{\{\hat{Q}_y(\tau_1\,|\,\bx_i) - \hat{Q}_y(\tau_2\,|\,\bx_i) \le 0 \} }  \neq n $}{
    Re-solve problem \eqref{eq:pcqr} using the updated $\gamma$\;
    }{
    $\mathrm{out} = 1$\;
    }
    $\gamma = \gamma + 0.01$\;
 }
 \KwResult{$\gamma^*$}
\caption{Searching the optimal tuning parameter $\gamma^*$.}
\end{algorithm}
\vspace{2em}

We proceed to illustrate how non-crossing quantile functions look like
with a real dataset used in \citeasnoun{Kuosmanen2020b}. It contains plant-level
data on 130  U.S. electric power plants operating in 2014; see
\citeasnoun{Kuosmanen2020b} for a more detailed description of the data. For the sake of demonstration, we simply consider a univariate case of one input and one output. The input is the total cost involved in electricity production and the output is the net electricity generation of each power plant. Both variables are in natural logarithm.

An application of CQR to the empirical data finds that the 15$^\text{th}$ quantile curve crosses the 25$^\text{th}$ quantile curve twice (see Fig.~\ref{fig1}). We then demonstrate how the sCQR and pCQR approaches can address this problem. It is evident from Fig.~\ref{fig2} that
both approaches manage to circumvent the quantile crossing problem; that is, we observe that $\widehat{Q}_y(0.25\,|\,\bx_i)$ is greater than or equal to
$\widehat{Q}_y(0.15\,|\,\bx_i)$ in both approaches. However, the shapes of the estimated quantile functions in Figs.~\ref{fig2:a} and \ref{fig2:b} (see particularly the upper right corner) are slightly different. As mentioned earlier, this difference arises because sCQR tries to simultaneously optimize for the non-crossing property, the quantile property, and the production axioms, whereas the pCQR approach independently estimates the quantile production functions. Further, the difference affects which approach can better retain the quantile property.

%

\section{Monte Carlo study}\label{sec:mc}

We perform a Monte Carlo study to examine whether pCQR or sCQR can better satisfy the quantile property while addressing quantile crossing. Consider the following data generating process (\citename{Dai2021a}, \citeyear*{Dai2021a}) 
\begin{equation*}
	y_i = \prod\limits_{d=1}^{D}\bx^{\frac{0.8}{d}}_{d,i} + v_i - u_i,
\end{equation*}
where the input matrix $\bx_i \in \real^{n \times d}$ is generated independently
from $U[1, 10]$, and noise $v_i$ and inefficiency $u_i$ are drawn independently
from $N(0, \sigma_v^2)$ and $N^+(0, \sigma_u^2)$, respectively. To investigate the robustness of the quantile approaches, following \citeasnoun{Aigner1977} we use different combinations of noise ($\sigma$) and signal to noise ratio ($\lambda$), that is, ($\sigma^2$, $\lambda$) = (1.88, 1.66), (1.63, 1.24), and (1.35, 0.83).

We consider 54 scenarios with $n \in \{99, 199, 499\}$, $d \in \{2, 3, 4\}$, and $\tau \in \{0.85, 0.90, 0.95\}$. Each scenario is replicated 500 times using the pyStoNED package (\citename{Dai2021b}, \citeyear*{Dai2021b}) on Python with the standard solver Mosek (9.3). We then compute the ramp loss ($\text{RL}=| \frac{1}{n} \sum_{i=1}^{n} \textbf{1}_{y_i > {{\widehat{Q}}_{y}}(\tau\,|\,\bx_i)} - \tau |$) (\citename{Takeuchi2006}, \citeyear*{Takeuchi2006}) to examine the quantile property and the mean squared error (MSE) to evaluate the finite-sample performance. Replications for which no quantile crossing happen (i.e., $\gamma^*=0$) are excluded from the calculations of RL and MSE. Note that the smaller the ramp loss, the better the quantile performance. 

\begin{figure}[H]
	\centering
	\includegraphics[width=1\textwidth]{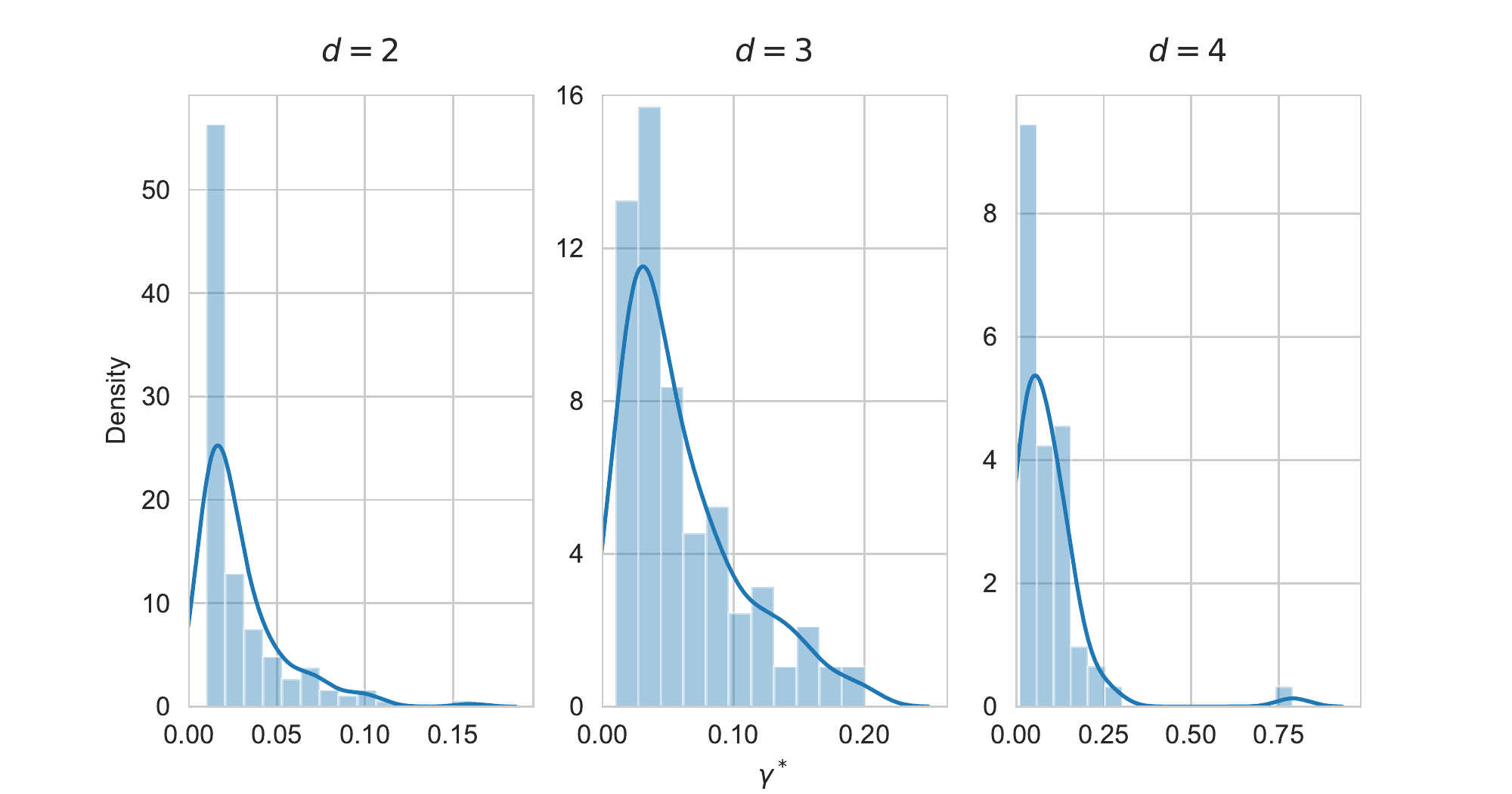} 
	\caption[]%
	{\small Empirical distribution of $\gamma^*$ with $n=499$, $(\sigma^2, \lambda) = (1.35, 0.83)$, $\tau_1=0.85$, and $\tau_2=0.90$.} 
	\label{fig3}
\end{figure}

Tables~\ref{tab:tab1} and \ref{tab:a1} present the estimated ramp loss and MSE statistics across different scenarios. The results clearly show that compared with sCQR, pCQR has lower ramp loss in virtually all the scenarios and lower MSE in all the scenarios. This is because the optimal tuning parameter $\gamma^*$ used in the pCQR simulations mainly locates in the interval $[0.01, 0.03]$ (see Fig.~\ref{fig3}), suggesting that the quantile property can be guaranteed to a certain extent and thus pCQR can better fit the true quantile functions. Several other findings are summarized as follows:
\begin{itemize}
    \item The higher the dimension or the noise in data space, the lower the ramp loss. As $d$ or $\sigma$ increases, the data space becomes more sparse, thereby indicating that the probability of crossing between two neighboring quantiles is relatively small. 
    \item The differences in the ramp loss among quantiles in pCQR are smaller than those in sCQR due to the different estimation strategies, i.e., independent and simultaneous estimation, respectively. 
    \item For both approaches, the MSE increases as more inputs are included and decreases as the sample size gets larger. 
    \item The performance of both approaches in terms of MSE becomes worse as the signal to noise ratio increases. 
\end{itemize}

Overall, the pCQR approach can better satisfy the quantile property and fit the
true quantile functions while at the same time addressing the quantile crossing
problem. Regularizing the quantile function, instead of imposing an extra set of non-crossing constraints, proves a better remedy to quantile crossing according to our simulations. 

\begin{table}[H]
  \centering
   \caption{Ramp loss and MSE of two neighboring quantiles with $n=499$.}
   \renewcommand\arraystretch{1.2}
   \footnotesize{
    \begin{tabular}{rrrrrrrrrrrr}
    \toprule
    \multicolumn{1}{c}{\multirow{2}[4]{*}{$d$}} & \multicolumn{1}{c}{\multirow{2}[4]{*}{($\sigma^2, \lambda$)}} & \multicolumn{1}{c}{\multirow{2}[4]{*}{$\tau_1$}} & \multicolumn{1}{c}{\multirow{2}[4]{*}{$\tau_2$}} & \multicolumn{2}{c}{RL$_{\tau_1}$} & \multicolumn{2}{c}{MSE$_{\tau_1}$} & \multicolumn{2}{c}{RL$_{\tau_2}$} & \multicolumn{2}{c}{MSE$_{\tau_2}$} \\
\cmidrule{5-12}          &       &       &       & \multicolumn{1}{c}{pCQR} & \multicolumn{1}{c}{sCQR} & \multicolumn{1}{c}{pCQR} & \multicolumn{1}{c}{sCQR} & \multicolumn{1}{c}{pCQR} & \multicolumn{1}{c}{sCQR} & \multicolumn{1}{c}{pCQR} & \multicolumn{1}{c}{sCQR} \\
    \midrule
    2     & \multicolumn{1}{l}{(1.35, 0.83)} & 0.85  & 0.90  & 0.928 & 0.993 & 0.065 & 0.093 & 0.968 & 1.035 & 0.075 & 0.115 \\
          &       & 0.90  & 0.95  & 0.998 & 1.078 & 0.074 & 0.117 & 1.035 & 1.108 & 0.095 & 0.176 \\
          & \multicolumn{1}{l}{(1.63, 1.24)} & 0.85  & 0.90  & 0.899 & 0.953 & 0.075 & 0.108 & 0.933 & 1.028 & 0.085 & 0.135 \\
          &       & 0.90  & 0.95  & 0.972 & 1.061 & 0.081 & 0.134 & 0.991 & 1.051 & 0.108 & 0.208 \\
          & \multicolumn{1}{l}{(1.88, 1.66)} & 0.85  & 0.90  & 0.898 & 0.967 & 0.079 & 0.117 & 0.984 & 1.069 & 0.095 & 0.155 \\
          &       & 0.90  & 0.95  & 0.890 & 0.963 & 0.093 & 0.153 & 0.923 & 1.012 & 0.117 & 0.228 \\
    3     & \multicolumn{1}{l}{(1.35, 0.83)} & 0.85  & 0.90  & 0.811 & 0.942 & 0.103 & 0.205 & 0.826 & 0.947 & 0.115 & 0.265 \\
          &       & 0.90  & 0.95  & 0.775 & 0.902 & 0.122 & 0.253 & 0.745 & 0.860 & 0.166 & 0.402 \\
          & \multicolumn{1}{l}{(1.63, 1.24)} & 0.85  & 0.90  & 0.832 & 0.950 & 0.110 & 0.226 & 0.832 & 0.940 & 0.127 & 0.296 \\
          &       & 0.90  & 0.95  & 0.827 & 0.959 & 0.138 & 0.281 & 0.751 & 0.878 & 0.175 & 0.435 \\
          & \multicolumn{1}{l}{(1.88, 1.66)} & 0.85  & 0.90  & 0.792 & 0.913 & 0.127 & 0.250 & 0.827 & 0.960 & 0.144 & 0.325 \\
          &       & 0.90  & 0.95  & 0.789 & 0.914 & 0.172 & 0.332 & 0.749 & 0.899 & 0.214 & 0.512 \\
    4     & \multicolumn{1}{l}{(1.35, 0.83)} & 0.85  & 0.90  & 0.658 & 0.837 & 0.163 & 0.368 & 0.671 & 0.833 & 0.187 & 0.484 \\
          &       & 0.90  & 0.95  & 0.627 & 0.794 & 0.207 & 0.487 & 0.481 & 0.601 & 0.269 & 0.750 \\
          & \multicolumn{1}{l}{(1.63, 1.24)} & 0.85  & 0.90  & 0.587 & 0.801 & 0.168 & 0.397 & 0.627 & 0.795 & 0.193 & 0.526 \\
          &       & 0.90  & 0.95  & 0.696 & 0.861 & 0.217 & 0.535 & 0.539 & 0.700 & 0.279 & 0.828 \\
          & \multicolumn{1}{l}{(1.88, 1.66)} & 0.85  & 0.90  & 0.584 & 0.793 & 0.189 & 0.446 & 0.622 & 0.793 & 0.218 & 0.589 \\
          &       & 0.90  & 0.95  & 0.646 & 0.795 & 0.285 & 0.581 & 0.504 & 0.640 & 0.381 & 0.920 \\
    \bottomrule
    \end{tabular}%
    }
	\label{tab:tab1}%
\end{table}%
\vspace{-1.5em}

%

\section{Conclusions}\label{sec:concl}

In this paper, a penalized convex quantile regression approach has been developed to address the quantile crossing problem. The proposed algorithm can search the optimal tuning parameter such that the occurrences of quantile crossing are avoided and the quantile property is ensured as well as possible. A Monte Carlo study confirms the superiority of the proposed approach compared to previous work in addressing the quantile crossing problem. We believe the proposed approach can be readily extended from convex quantile regression to other nonparametric and parametric quantile regression techniques. We leave such extensions as fascinating avenues for future research.

%

\baselineskip 12pt
\bibliography{references}

%

\newpage
\section*{Appendix}\label{sec:app}

\renewcommand{\thesubsection}{\Alph{subsection}}
\renewcommand{\thetable}{A\arabic{table}}
\setcounter{table}{0}
\renewcommand{\thefigure} {A\arabic{figure}}
\setcounter{figure}{0}

\subsection{Supplementary tables and figures} \label{App:qcrossing}

\begin{figure}[H]
	\centering
	\caption{Illustration of the quantile crossing problem in CQR estimation.}
	\scalebox{1.0}{\includegraphics[width=5.9truein]{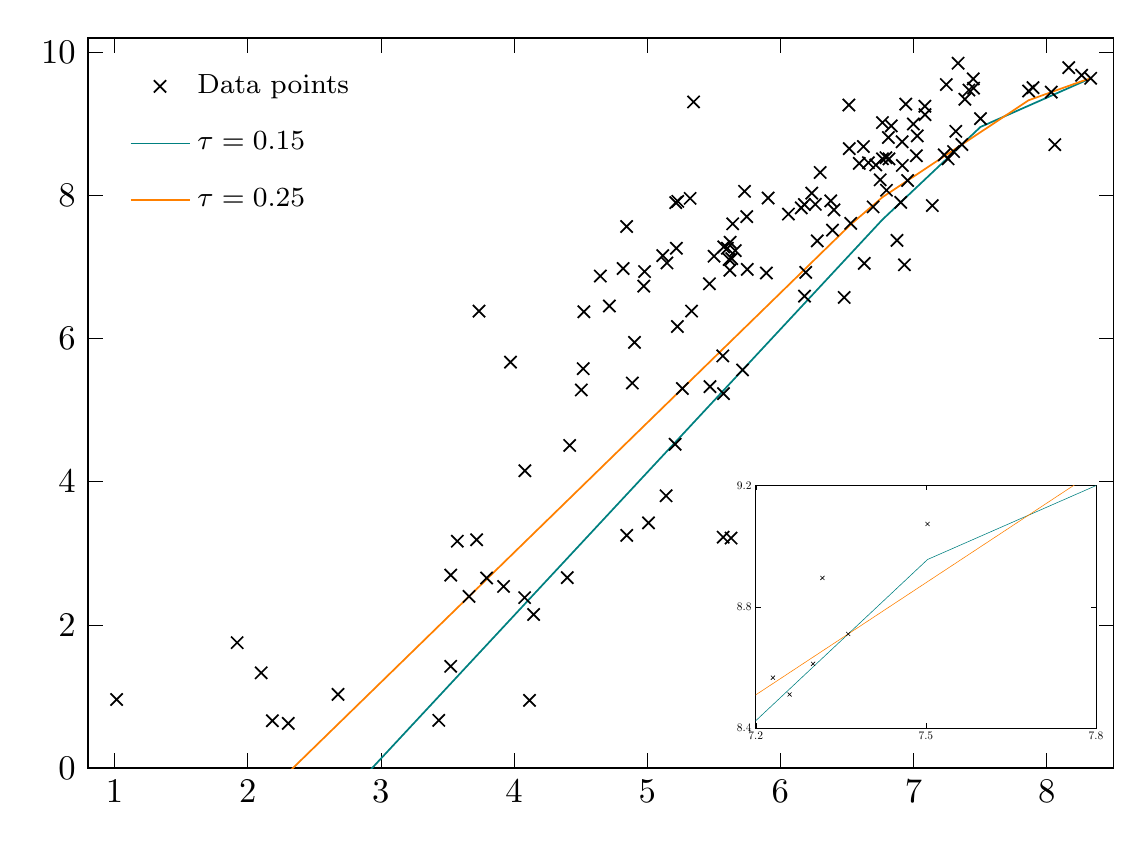}}
	\label{fig1}
\end{figure}

\newpage
\begin{table}[htbp]
  \centering
  	\caption{Ramp loss and MSE of two neighboring quantiles with $n=99$ and $199$.}
    \renewcommand\arraystretch{1.2}
    \footnotesize{
    \begin{tabular}{rrrccrrrrrrrr}
    \toprule
    \multicolumn{1}{c}{\multirow{2}[4]{*}{$n$}} & \multicolumn{1}{c}{\multirow{2}[4]{*}{$d$}} & \multicolumn{1}{c}{\multirow{2}[4]{*}{($\sigma^2, \lambda$)}} & \multirow{2}[4]{*}{$\tau_1$} & \multirow{2}[4]{*}{$\tau_2$} & \multicolumn{2}{c}{RL$_{\tau_1}$} & \multicolumn{2}{c}{MSE$_{\tau_1}$} & \multicolumn{2}{c}{RL$_{\tau_2}$} & \multicolumn{2}{c}{MSE$_{\tau_2}$} \\
\cmidrule{6-13}          &       &       &       &       & \multicolumn{1}{c}{pCQR} & \multicolumn{1}{c}{sCQR} & \multicolumn{1}{c}{pCQR} & \multicolumn{1}{c}{sCQR} & \multicolumn{1}{c}{pCQR} & \multicolumn{1}{c}{sCQR} & \multicolumn{1}{c}{pCQR} & \multicolumn{1}{c}{sCQR} \\
    \midrule
99    & 2     & \multicolumn{1}{l}{(1.35, 0.83)} & 0.85  & 0.90  & 0.824 & 0.882 & 0.239 & 0.350 & 0.869 & 0.884 & 0.276 & 0.444 \\
&       &       & 0.90  & 0.95  & 0.849 & 0.899 & 0.300 & 0.459 & 0.835 & 0.794 & 0.396 & 0.683 \\
&       & \multicolumn{1}{l}{(1.63, 1.24)} & 0.85  & 0.90  & 0.806 & 0.882 & 0.248 & 0.378 & 0.857 & 0.904 & 0.287 & 0.484 \\
&       &       & 0.90  & 0.95  & 0.888 & 0.935 & 0.324 & 0.481 & 0.867 & 0.845 & 0.427 & 0.727 \\
&       & \multicolumn{1}{l}{(1.88, 1.66)} & 0.85  & 0.90  & 0.831 & 0.898 & 0.272 & 0.435 & 0.891 & 0.916 & 0.327 & 0.556 \\
&       &       & 0.90  & 0.95  & 0.857 & 0.918 & 0.336 & 0.534 & 0.792 & 0.809 & 0.439 & 0.803 \\
& 3     & \multicolumn{1}{l}{(1.35, 0.83)} & 0.85  & 0.90  & 0.719 & 0.753 & 0.349 & 0.619 & 0.667 & 0.702 & 0.400 & 0.777 \\
&       &       & 0.90  & 0.95  & 0.626 & 0.687 & 0.442 & 0.807 & 0.446 & 0.381 & 0.618 & 1.262 \\
&       & \multicolumn{1}{l}{(1.63, 1.24)} & 0.85  & 0.90  & 0.671 & 0.721 & 0.400 & 0.688 & 0.632 & 0.657 & 0.469 & 0.878 \\
&       &       & 0.90  & 0.95  & 0.629 & 0.695 & 0.516 & 0.875 & 0.393 & 0.364 & 0.713 & 1.392 \\
&       & \multicolumn{1}{l}{(1.88, 1.66)} & 0.85  & 0.90  & 0.677 & 0.741 & 0.439 & 0.782 & 0.622 & 0.676 & 0.500 & 0.989 \\
&       &       & 0.90  & 0.95  & 0.611 & 0.662 & 0.541 & 1.046 & 0.437 & 0.404 & 0.742 & 1.651 \\
& 4     & \multicolumn{1}{l}{(1.35, 0.83)} & 0.85  & 0.90  & 0.412 & 0.491 & 0.591 & 1.041 & 0.334 & 0.342 & 0.715 & 1.384 \\
&       &       & 0.90  & 0.95  & 0.382 & 0.421 & 0.798 & 1.381 & 0.239 & 0.194 & 1.118 & 2.148 \\
&       & \multicolumn{1}{l}{(1.63, 1.24)} & 0.85  & 0.90  & 0.455 & 0.550 & 0.688 & 1.139 & 0.403 & 0.427 & 0.833 & 1.497 \\
&       &       & 0.90  & 0.95  & 0.381 & 0.395 & 0.874 & 1.491 & 0.229 & 0.182 & 1.264 & 2.384 \\
&       & \multicolumn{1}{l}{(1.88, 1.66)} & 0.85  & 0.90  & 0.502 & 0.606 & 0.664 & 1.180 & 0.450 & 0.468 & 0.813 & 1.567 \\
&       &       & 0.90  & 0.95  & 0.412 & 0.449 & 0.931 & 1.624 & 0.239 & 0.231 & 1.339 & 2.567 \\
& \vspace{-0.5em} \\
199   & 2     & \multicolumn{1}{l}{(1.35, 0.83)} & 0.85  & 0.90  & 0.837 & 0.916 & 0.139 & 0.201 & 0.914 & 0.964 & 0.160 & 0.251 \\
&       &       & 0.90  & 0.95  & 0.914 & 0.997 & 0.159 & 0.250 & 0.934 & 0.981 & 0.206 & 0.372 \\
&       & \multicolumn{1}{l}{(1.63, 1.24)} & 0.85  & 0.90  & 0.869 & 0.957 & 0.140 & 0.210 & 0.935 & 0.978 & 0.166 & 0.276 \\
&       &       & 0.90  & 0.95  & 0.944 & 1.016 & 0.178 & 0.291 & 0.959 & 1.003 & 0.233 & 0.427 \\
&       & \multicolumn{1}{l}{(1.88, 1.66)} & 0.85  & 0.90  & 0.868 & 0.931 & 0.150 & 0.236 & 0.894 & 0.949 & 0.186 & 0.311 \\
&       &       & 0.90  & 0.95  & 0.945 & 1.012 & 0.186 & 0.322 & 0.948 & 0.968 & 0.223 & 0.459 \\
& 3     & \multicolumn{1}{l}{(1.35, 0.83)} & 0.85  & 0.90  & 0.656 & 0.777 & 0.203 & 0.374 & 0.703 & 0.779 & 0.231 & 0.485 \\
&       &       & 0.90  & 0.95  & 0.733 & 0.830 & 0.256 & 0.472 & 0.561 & 0.621 & 0.338 & 0.721 \\
&       & \multicolumn{1}{l}{(1.63, 1.24)} & 0.85  & 0.90  & 0.696 & 0.815 & 0.221 & 0.436 & 0.708 & 0.790 & 0.248 & 0.568 \\
&       &       & 0.90  & 0.95  & 0.748 & 0.881 & 0.266 & 0.550 & 0.631 & 0.707 & 0.364 & 0.848 \\
&       & \multicolumn{1}{l}{(1.88, 1.66)} & 0.85  & 0.90  & 0.713 & 0.837 & 0.249 & 0.461 & 0.740 & 0.839 & 0.277 & 0.595 \\
&       &       & 0.90  & 0.95  & 0.657 & 0.716 & 0.335 & 0.653 & 0.516 & 0.553 & 0.438 & 1.013 \\
& 4     & \multicolumn{1}{l}{(1.35, 0.83)} & 0.85  & 0.90  & 0.584 & 0.703 & 0.282 & 0.663 & 0.568 & 0.597 & 0.328 & 0.876 \\
&       &       & 0.90  & 0.95  & 0.452 & 0.570 & 0.409 & 0.918 & 0.350 & 0.364 & 0.537 & 1.430 \\
&       & \multicolumn{1}{l}{(1.63, 1.24)} & 0.85  & 0.90  & 0.503 & 0.670 & 0.331 & 0.736 & 0.507 & 0.605 & 0.380 & 0.969 \\
&       &       & 0.90  & 0.95  & 0.453 & 0.569 & 0.476 & 0.974 & 0.321 & 0.354 & 0.617 & 1.533 \\
&       & \multicolumn{1}{l}{(1.88, 1.66)} & 0.85  & 0.90  & 0.428 & 0.622 & 0.354 & 0.806 & 0.427 & 0.554 & 0.423 & 1.074 \\
&       &       & 0.90  & 0.95  & 0.472 & 0.630 & 0.515 & 1.084 & 0.332 & 0.375 & 0.657 & 1.687 \\
\bottomrule
    \end{tabular}%
    }
	\label{tab:a1}%
\end{table}%

\end{document}